\documentclass[10pt,preprint2]{aastex}

\newcommand{\Mks}{\ifmmode M_{K_{\rm s}} \else $M_{K_{\rm s}}$ \fi} 
\newcommand{\ff}{\ifmmode f_{{\rm gm}} \else $f_{{\rm gm}}$ \fi}
\newcommand{\ffo}{\ifmmode f_{{\rm gm},0} \else $f_{{\rm gm},0}$ \fi}
\newcommand{\ffk}{\ifmmode f_{{\rm gm},k} \else $f_{{\rm gm},k}$ \fi}
\newcommand{\ffkr}{\ifmmode f_{{\rm gm},k}^{\rm in} \else $f_{{\rm gm},k}^{\rm in}$ \fi}
\newcommand{\ffkml}{\ifmmode f_{{\rm gm},k}^{\rm ML} \else $f_{{\rm gm},k}^{\rm ML}$ \fi}
\newcommand{\ffkclass}{\ifmmode f_{{\rm gm},k}^{\rm class} \else $f_{{\rm gm},k}^{\rm class}$ \fi}
\newcommand{\sffk}{\ifmmode \sigma_{f_{{\rm gm},k}} \else $\sigma_{f_{{\rm gm},k}}$ \fi}
\newcommand{\sffkml}{\ifmmode \sigma_{f_{{\rm gm},k}^{\rm ML}} \else $\sigma_{f_{{\rm gm},k}^{\rm ML}}$ \fi}
\newcommand{\sffkmlp}{\ifmmode \sigma^{+}_{f_{{\rm gm},k}^{\rm ML}} \else $\sigma^{+}_{f_{{\rm gm},k}^{\rm ML}}$ \fi}
\newcommand{\sffkmlm}{\ifmmode \sigma^{-}_{f_{{\rm gm},k}^{\rm ML}} \else $\sigma^{-}_{f_{{\rm gm},k}^{\rm ML}}$ \fi}
\newcommand{\sffkclass}{\ifmmode \sigma_{f_{{\rm gm},k}^{\rm class}} \else $\sigma_{f_{{\rm gm},k}^{\rm class}}$ \fi}

\newcommand{\zp}{\ifmmode z_{{\rm phot}} \else $z_{{\rm phot}}$ \fi}
\newcommand{\zspec}{\ifmmode z_{{\rm spec}} \else $z_{{\rm spec}}$ \fi}

\newcommand{\pkl}{\ifmmode p_{kl} \else $p_{kl}$ \fi}
\newcommand{\pklp}{\ifmmode p'_{kl} \else $p'_{kl}$ \fi}
\newcommand{\spkl}{\ifmmode \sigma_{p_{kl}} \else $\sigma_{p_{kl}}$ \fi}
\newcommand{\spklp}{\ifmmode \sigma_{p'_{kl}} \else $\sigma_{p'_{kl}}$ \fi}

\newcommand{\pklrc}{\ifmmode p_{kl,{\rm ML}} \else $p_{kl,{\rm ML}}$ \fi}
\newcommand{\pklrcp}{\ifmmode p'_{kl,{\rm ML}} \else $p'_{kl,{\rm ML}}$ \fi}
\newcommand{\pklr}{\ifmmode p_{kl, {\rm in}} \else $p_{kl, {\rm in}}$ \fi}
\newcommand{\pklrp}{\ifmmode p'_{kl, {\rm in}} \else $p'_{kl, {\rm in}}$ \fi}
\newcommand{\mpklrc}{\ifmmode \overline{p_{kl, {\rm ML}}} \else $\overline{p_{kl, {\rm ML}}}$ \fi}
\newcommand{\mpklrcp}{\ifmmode \overline{p'_{kl, {\rm ML}}} \else $\overline{p'_{kl, {\rm ML}}}$ \fi}

\newcommand{\spklrc}{\ifmmode \sigma_{p_{kl,{\rm ML}}} \else $\sigma_{p_{kl,{\rm ML}}}$ \fi}
\newcommand{\spklrcp}{\ifmmode \sigma_{p'_{kl,{\rm ML}}} \else $\sigma_{p'_{kl,{\rm ML}}}$ \fi}
\newcommand{\smpklrc}{\ifmmode \sigma_{\overline{p_{kl,{\rm ML}}}} \else $\sigma_{\overline{p_{kl,{\rm ML}}}}$ \fi}
\newcommand{\smpklrcp}{\ifmmode s_{p'_{kl,{\rm ML}}} \else $s_{p'_{kl,{\rm ML}}}$ \fi}
\newcommand{\mspklrc}{\ifmmode \overline{\sigma_{p_{kl,{\rm ML}}}} \else $\overline{\sigma_{p_{kl,{\rm ML}}}}$ \fi}
\newcommand{\mspklrcp}{\ifmmode \overline{\sigma_{p'_{kl,{\rm ML}}}} \else $\overline{\sigma_{p'_{kl,{\rm ML}}}}$ \fi}
\newcommand{\spklML}{\ifmmode \sigma_{p_{kl,{\rm iter}}} \else $\sigma_{p_{kl,{\rm iter}}}$ \fi}
\newcommand{\spklMLp}{\ifmmode s_{p'_{kl,{\rm iter}}} \else $s_{p'_{kl,{\rm iter}}}$ \fi}
\newcommand{\covpklrcp}{\ifmmode {\rm cov}(p'_{mn,{\rm ML}},p'_{st,{\rm ML}}) \else ${\rm cov}(p'_{mn,{\rm ML}},p'_{st,{\rm ML}})$ \fi}

\newcommand{\pklclass}{\ifmmode p_{kl, {\rm class}} \else p_{kl, {\rm class}}$ \fi}
\newcommand{\pklclassp}{\ifmmode p'_{kl, {\rm class}} \else p'_{kl, {\rm class}}$ \fi}
\newcommand{\mpklclass}{\ifmmode \overline{p_{kl, {\rm class}}} \else $\overline{p_{kl, {\rm class}}}$ \fi}
\newcommand{\mpklclassp}{\ifmmode \overline{p'_{kl, {\rm class}}} \else $\overline{p'_{kl, {\rm class}}}$ \fi}
\newcommand{\smpklclass}{\ifmmode s_{p_{kl, {\rm class}}} \else $s_{p_{kl, {\rm class}}}$ \fi}
\newcommand{\smpklclassp}{\ifmmode s_{p'_{kl, {\rm class}}} \else $s_{p'_{kl, {\rm class}}}$ \fi}
\newcommand{\mspklclass}{\ifmmode \overline{\sqrt{p_{kl, {\rm class}}}} \else $\overline{\sqrt{p_{kl, {\rm class}}}}$ \fi}
\newcommand{\mspklclassp}{\ifmmode \overline{p'_{kl, {\rm theo}}}  \else $\overline{p'_{kl, {\rm theo}}}$ \fi}
\newcommand{\spklclassp}{\ifmmode p'_{kl, {\rm theo}} \else $p'_{kl, {\rm theo}}$ \fi}

\shorttitle{Galaxy merger fractions using maximum likelihood techniques}
\shortauthors{L\'opez-Sanjuan et al.}

\begin{document}

\title{A maximum likelihood method for bidimensional experimental distributions, and its application to the galaxy merger fraction}
\author{Carlos L\'opez-Sanjuan, C\'esar Enrique Garc\'{\i}a-Dab\'o, Marc Balcells}
\affil{Instituto de Astrof\'{\i}sica de Canarias, Calle V\'{\i}a L\'actea s/n, La Laguna, Tenerife, 38200 Spain}
\email{clsj@iac.es, enrique.garcia@gtc.iac.es, balcells@iac.es}

\begin{abstract}
The determination of galaxy merger fraction of field galaxies using automatic morphological indices and photometric redshifts is affected by several biases if observational errors are not properly treated. Here, we correct these biases using maximum likelihood techniques. The method takes into account the observational errors to statistically recover the real shape of the bidimensional distribution of galaxies in redshift - asymmetry space, needed to infer the redshift evolution of galaxy merger fraction. We test the method with synthetic catalogs and show its applicability limits. The accuracy of the method depends on catalog characteristics such as the number of sources or the experimental error sizes.  We show that the maximum likelihood method recovers the real distribution of galaxies in redshift and asymmetry space even when binning is such that bin sizes approach the size of the observational errors. We provide a step-by-step guide to applying maximum likelihood techniques to recover any one- or bidimensional distribution subject to observational errors.  
\end{abstract}

\keywords{Data Analysis and Techniques}

\section{INTRODUCTION}

The currently popular hierarchical $\Lambda$CDM models are successful at explaining the structure build-up of the cold dark matter component of the Universe \citep{springel05}. But such models have difficulties when explaining the evolution of the baryonic component, even with modeling that incorporates star formation, active galactic nuclei and supernova feedback, and the multiphase nature of the interstellar medium  \citep[][and references therein]{delucia07}. An open question is the role of the galaxy mergers in the formation of today's galaxies, specially the most massive ellipticals. The observational determination of the merger rate, $\Re_{{\rm m}}$, and its evolution with redshift, provide empirical clues on the amount and the timing of the merger activity.  They also constitute key inputs for semi-analytic models of galaxy formation and evolution.  

The merger rate, defined as the number density of merger systems at given redshift, depends on the merger time $\tau_{m}$, which can only be estimated by N-body simulations and simplified models \citep[]{mihos95,patton00,conselice06}. On the other hand, the galaxy merger fraction $\ff$, defined as the number of merger galaxies in a given galaxy sample in a redshift interval, is a direct observational quantity. Many works have determined the galaxy merger fraction, usually parametrized as $\ff = \ffo \cdot (1+z)^m$, using different sample selection and methods, like morphological criteria \citep[]{conselice03, lavery04, cassata05, lotz06, bridge07, depropris07}, kinematic close companions \citep[]{patton00, patton02, lin04, depropris05, depropris07}, spatial close pairs \citep[]{lefevre00, bundy04, bridge07, kartaltepe07} or correlation function \citep[]{bell06, masjedi06}. In these works the value of the merger index varies in the range $m = 0 - 4$. $\Lambda$CDM models predict $m \sim 3$ \citep[]{kolatt99, governato99, gottlober01}.

The morphological criterion for determining the galaxy merger fraction \citep[see][hereafter C03]{conselice03}, is based on the fact that, just after a merger is complete, the galaxy image shows strong geometrical distortions, in particular asymmetric distortions. Hence, high values in the automatic asymmetry index $A$ (\citealt{abraham96}; C03) are assumed to identify merger systems. Other automatic morphological indices, such as $M_{20}$ and $G$, have also been used to determine the evolution of galaxy merger fraction with redshift \citep[]{lotz06}. The determination of morphological indices, which must be done on \facility{HST} images, is affected by surface brightness dimming and K-corrections, so the errors of the indices grow with redshift and are more important for faint galaxies.  

In this paper, we present a method based on the maximum likelihood (ML) technique, to handle the effects of the large errors on the determination of the galaxy merger fraction. Galaxy Merger fraction determinations using morphological criteria are generally done on large photometric surveys such as AEGIS \citep{davis07}, COMBO-17 \citep{wolf03}, COSMOS \citep{scoville07}, GOODS \citep{giavalisco04}, or SWIRE \citep{lonsdale03}.  We therefore address the effects of errors in the galaxy asymmetry indices as well as errors on the photometric redshifts.  

In Section \ref{methodology} we review the maximum likelihood method for determining bidimensional distributions. Its application to the galaxy merger fraction determination is given in Section \ref{teofmg}. These sections have a high mathematical content, and a statistics background is recommended. Then, in Section \ref{simulations} we summarize the simulations made to test the general method and how it improves the galaxy merger fraction determination, Section \ref{FmgComp}. In Section \ref{MLsteps} we provide an outline for the application of the ML method to any one- or bidimensional experimental distribution subject to observational errors. Our conclusions are presented in Section \ref{conclusions}.

\section{METHODOLOGY}\label{methodology}
Following \cite{conselice06}, we define the galaxy merger fraction by morphological criteria as
\begin{equation}\label{fmg}
\ff = \frac{\kappa \cdot N_{\rm m}}{N_{\rm tot} + (\kappa -1) N_{\rm m}},
\end{equation}
where $N_{\rm m}$ is the number of the distorted sources in the sample, classified as the systems with a value in the asymmetry index $A$ higher than a limiting value $A_{\rm m}$ (see C03 for details), $N_{\rm tot}$ is the total number of sources in the sample, and $\kappa$ is the average number of galaxies that merged to produce the $N_{\rm m}$ merger systems. We use $\kappa = 2$ throughout this paper.

In order to compute the galaxy merger fraction and its redshift evolution we must know the underlying distribution of the $z$ and $A$ values, that we assume is represented by a bidimensional histogram in redshift and asymmetry space. This bidimensional histogram is defined by the number of sources in each redshift-asymmetry bin. Normalizing to unity the histogram yields a bidimensional probability distribution defined now by $p_{kl}$, the probability that a source has redshift in bin $k$ and asymmetry in bin $l$. Index $k$ scans the redshift bins of size $\Delta z$ and index $l$ scans the asymmetry bins of size $\Delta A$. In our case we just need two asymmetry bins separated by $A_{\rm m}$: the $l=0$ bin represents normal sources and the $l=1$ bin represents merger systems. Now, the galaxy merger fraction in redshift bin $[z_{k}, z_{k+1})$ is
\begin{equation}\label{fmgf}
\ffk = \frac{2p_{k1}}{p_{k0}+2p_{k1}}.
\end{equation}

The accuracy with which the $p_{kl}$ can be obtained degrades significantly when photometric redshifts, $\zp$, are used, and for typical errors of $A$ in deep \facility{HST} surveys. This introduces strong biases in the determination of the galaxy merger fraction.

\subsection{The maximum likelihood method}
The maximum likelihood method (ML method) developed here is based on \citet{garciadabo02}, who used this technique to determine unbiased luminosity functions. ML methods have been used in a wide range of topics in astrophysics.  \citet{arzner07} use it to improve the determination of faint X-ray spectra; \citet{sheth07} to obtain redshift and luminosity distributions in photometric surveys; \citet{naylor06} to fit colour-magnitude diagrams; \citet{makarov06} to improve distance estimates using Red Giant Branch stars; and, \citet{efstathiou04} to analyze low cosmic microwave background multipoles from the Wilkinson Microwave Anisotropy Probe. ML methods are based on the estimation of the most probable values of a set of parameters which define the probability distribution that describes an observational sample \citep{davidson93,penha01}. 

The general ML method operates as follows. Throughout the paper we denote as ${\it P}({\bf a}|{\bf b})$ the probability to obtain the values ${\bf a}$, given parameters ${\bf b}$. Being ${\bf{x}}_i$ a vector containing all the measured values for source $i$ in the data set and $\theta$ the parameters of the underlying multidimensional distribution that we want to estimate, we may express the joined likelihood function as
\begin{equation}\label{MLdef}
L({\bf x}_i| \theta ) \equiv -\ln \big[ \prod_i {\it P}( {\bf x}_i | \theta) \big] = - \sum_i \ln \big[{\it P}({\bf x}_i|{\bf \theta})\big],
\end{equation}
where ${\it P}({\bf x}_i|\theta)$ is the probability to obtain ${{\bf x}_i}$ for a given $\theta$. If we are able to express ${\it P}({\bf x}_i|\theta)$ analytically, we can minimize Equation \ref{MLdef} to obtain the best estimation of parameters $\theta$, denote as $\theta_{ML}$. In our case, ${\bf x}_i$ are the observed values of $z$ and $A$ for source $i$, ${\bf x}_i \equiv (z_ {{\rm obs},i}, A_{{\rm obs},i})$, while $\theta \equiv (p_{kl},\alpha)$, where $p_{kl}$ are the probabilities which we defined in the paragraph previous to Equation \ref{fmgf}, and $\alpha$ denotes any other fixed parameters of the distribution.

Sources are assumed to have real redshift and asymmetry values $z_{{\rm real},i}$ and $A_{{\rm real},i}$ (not affected by observational errors) which define a bidimensional distribution $p_{kl}$ such that

\begin{displaymath}
P_{2D}(z_{{\rm real},i},A_{{\rm real},i}|p_{kl})\ \ \ \ \ \ \ \ \ \ \ \ \ \ \ \ \ \ \ \ \ \ \ \ \ \ \ \ \ 
\end{displaymath}
\begin{equation}
= \{p_{kl} ,\forall z_k \leq z_{{\rm real},i} < z_{k+1} , A_l \leq A_{{\rm real},i} < A_{l+1}\}.\label{bidi}
\end{equation}

Observational errors cause the observed $z_{{\rm obs},i}$ and $A_{{\rm obs},i}$ to differ from their respective real values  $z_{{\rm real},i}$ and $A_{{\rm real},i}$.  
The observed $z_{{\rm obs},i}$ are assumed to be extracted for a Gaussian distribution with mean $z_{{\rm real},i}$ and standard deviation $\sigma_{z_{{\rm obs},i}}$, 
\begin{eqnarray}\label{zgauss}
\lefteqn{P_G(z_{{\rm obs},i}|z_{{\rm real},i},\sigma_{z_{{\rm obs},i}})}\nonumber\\ 
&& \ \ \ \ \ = \frac{1}{\sqrt{2\pi}\sigma_{z_i}}{\rm e}^{-\frac{(z_{{\rm obs},i}-z_{{\rm real},i})^2}{2\sigma_{z_{{\rm obs},i}}^2}}.
\end{eqnarray}

\noindent Similarly, the observed asymmetry values $A_{{\rm obs},i}$ are assumed to be extracted from a Gaussian distribution with mean $A_{{\rm real},i}$ and standard deviation $\sigma_{A_{{\rm obs},i}}$, 
\begin{eqnarray}\label{agauss}
\lefteqn{P_G(A_{{\rm obs},i}|A_{{\rm real},i},\sigma_{A_{{\rm obs},i}})} \nonumber\\ 
&& = \frac{1}{\sqrt{2\pi}\sigma_{A_{{\rm obs},i}}}{\rm e}^{-\frac{(A_{{\rm obs},i}-A_{{\rm real},i})^2}{2\sigma_{A_{{\rm obs},i}}^2}}.
\end{eqnarray}

While the $\zp$ errors may not be strictly Gaussian, this is the best analytical approximation of the errors that we can make. We obtain the probability ${\it P}({\bf x}_i| \theta )$ of each source by the total probability theorem:
\begin{eqnarray}\label{tprob}
\lefteqn{P(z_{{\rm obs},i},A_{{\rm obs},i}|p_{kl},\sigma_{z_{{\rm obs},i}},\sigma_{A_{{\rm obs},i}})} \nonumber\\
&& = \int P_G(z_{{\rm obs},i}|z_{{\rm real},i},\sigma_{z_{{\rm obs},i}})\nonumber\\
&& \times P_G(A_{{\rm obs},i}|A_{{\rm real},i},\sigma_{A_{{\rm obs},i}})\nonumber\\
&& \times P_{2D}(z_{{\rm real},i},A_{{\rm real},i}|p_{kl}){\rm d}z_{{\rm real},i}{\rm d}A_{{\rm real},i},
\end{eqnarray}
where ${{\bf x}_i} \equiv (z_{{\rm obs},i},A_{{\rm obs},i})$ and $\theta$ $\equiv (p_{kl},$ $\sigma_{z_{{\rm obs},i}},$ $\sigma_{A_{{\rm obs},i}})$ in Equation \ref{MLdef}, with $\alpha \equiv (\sigma_{z_{{\rm obs},i}},\sigma_{A_{{\rm obs},i}})$. Note that the values of $\sigma_{z_{{\rm obs},i}}$ and $\sigma_{A_{{\rm obs},i}}$ are the measured uncertainties for each source, so the only unknowns are the probabilities $p_{kl}$, which we want to estimate. Note also that we integrate over the variables $z_{{\rm real},i}$ and $A_{{\rm real},i}$, so we are not be able to estimate them individually, but only the underlying bidimensional distribution $p_{kl}$ that describes the sample.

In order to ensure that the probabilities \pkl are not negative, we change variables, $\pkl = {\rm exp}(\pkl')$; this change keeps our problem analytic. With these new variables and after integrating Equation \ref{tprob}, our likelihood function, defined in Equation \ref{MLdef}, becomes
\begin{displaymath}
L(z_{{\rm obs},i},A_{{\rm obs},i}|p^{\prime}_{kl},\sigma_{z_{{\rm obs},i}},\sigma_{A_{{\rm obs},i}})\ \ \ \ \ \ \ \ \ \ \ \ \ \ \ 
\end{displaymath}
\begin{equation}
= \sum_i \biggr[ \ln \bigg\{ \sum_k\sum_l \frac{{\rm e}^{p'_{kl}}}{4}{\rm ERF}(z,i,k){\rm ERF}(A,i,l) \bigg\}\biggr]\label{MLfunc},
\end{equation}
where
\begin{displaymath}
{\rm ERF}(\eta,i,k)\ \ \ \ \ \ \ \ \ \ \ \ \ \ \ \ \ \ \ \ \ \ \ \ \ \ \ \ \ \ \ \ \ \ \ \ \ \ 
\end{displaymath}
\begin{equation}
= {\rm erf}\bigg(\frac{\eta_{{\rm obs},i} - \eta_{k+1}}{\sqrt{2} \sigma_{\eta_{{\rm obs},i}}}\bigg) - {\rm erf}\bigg(\frac{\eta_{{\rm obs},i} - \eta_{k}}{\sqrt{2} \sigma_{\eta_{{\rm obs},i}}}\bigg),\label{ERF}
\end{equation}
and ${\rm erf}(x)$ is the error function. We must observe that in the minimization of Equation \ref{MLfunc} the variables $\pklp$ are not independent. This is due to the normalization of the distribition: the integration over all parameters space muts be one. This impose the following condition over $\pklp$: 
\begin{equation}\label{norm}
{\rm \bf{g}}(\pklp) \equiv \sum_k\sum_l  {\rm e}^{p'_{kl}} (z_{k+1} - z_{k}) (A_{l+1} - A_{l}) - 1 = 0.
\end{equation}
The method for finding the extrema of a function of several variables subject to one or more constraints is know as the Lagrange multipliers \citep[see e.g.,][for details]{marsden96}. It states that the function to minimize is not the target function, Equation \ref{MLfunc}, but a related one:
\begin{eqnarray}\label{lagrange}
G(\pklp,\lambda) =  L(z_{{\rm obs},i},A_{{\rm obs},i}|p^{\prime}_{kl},\sigma_{z_{{\rm obs},i}},\sigma_{A_{{\rm obs},i}})\nonumber\\
+ \lambda {\rm \bf{g}}(\pklp),
\end{eqnarray}
where $\lambda$ is an auxiliary variable. Minimizing Equation \ref{lagrange} we obtain the best \pklp values, denoted as $\pklrcp$.

The minimization of Equation \ref{lagrange} can be performed with any numerical minimization code. We used \texttt{AMOEBA}, which is based on the commonly used algorithm of Nelder-Mead \citep{nelder65} and coded in C \cite[][pp. 408-412]{press95}.

At this point we have the probabilities $\pklrcp$. However, our goal is to obtain not only the best probabilities estimation, but also their associated uncertainties. The ML method states that we can obtain all the parameter covariaces using an expansion of the function $G(\pklp,\lambda)$ in Taylor's series of its variables $\theta = (\pklp,\lambda)$ around the minimization point $\theta_{ML} = (\pklrcp,\lambda_{ML})$ if the probability distributions of \pklrcp are Gaussian, which we assume. The previous minimization process made the first $G$ derivate null at $\theta = \theta_{ML}$ and we obtain
\begin{equation}\label{taylor}
G = G(\theta_{ML}) + \frac{1}{2}(\theta - \theta_{ML})^T H (\theta - \theta_{ML}),
\end{equation}
where $H = h_{xy}$ is the Hessian matrix and $T$ denotes the transpose vector. The inverse of the Hessian matrix gives us an estimate of the 68\% confidence intervals of $\pklrcp$, denoted as $[\pklrcp - \spklrcp, \pklrcp - \spklrcp]$, and the covariances between each $\pklrcp$, denoted as $\covpklrcp$, because maximum likelihood theory states that ${\rm cov}(\theta_{x},\theta_{y}) \geq h^{-1}_{xy}$ and $\sigma_{\theta_{x}} \geq h^{-1}_{xx}$. In our case, the Hessian matrix is
\begin{equation}\label{hessian}
H = \left(
\begin{array}{cc}
  \frac{\partial^2 G}{\partial p'_{mn} \partial p'_{st}} & \triangledown g \\
  \triangledown g & 0 
\end{array}\right),
\end{equation}
where
\begin{eqnarray}
\frac{\partial^2 G}{\partial p'_{mn} \partial p'_{st}} = -\sum_i \frac{{\rm ERF}(z,i,m){\rm ERF}(A,i,n)}{16} \nonumber\\
\times \frac{{\rm ERF}(z,i,s){\rm ERF}(A,i,t){\rm e}^{p'_{mn}}{\rm e}^{p'_{st}}}{\sum_l\sum_k \frac{{\rm e}^{p'_{kl}}}{4}{\rm ERF}(z,i,k){\rm ERF}(A,i,l)}\label{sigmap}
\end{eqnarray}
\begin{equation}\label{sigmag}
\triangledown g = \frac{\partial^2 G}{\partial \lambda \partial p'_{mn}} = (z_{m+1} - z_{m}) (A_{n+1} - A_{n}){\rm e}^{p'_{mn}}.
\end{equation}

Finally, the $\pklrc$ probabilities simply are:
\begin{equation}\label{pML}
\pklrc = {\rm e}^{\pklrcp}.
\end{equation}
Assuming that the $\pklrcp$ follow a Gaussian distribution, which is assured by the ML theory for large number of sources, the $\pklrc$ follow a log-normal distribution:
\begin{eqnarray}
\lefteqn{P_{LN}(\pkl|\pklrcp,\spklrcp)} \nonumber\\ 
&& = \frac{{\rm e}^{-(\ln \pkl - \pklrcp)^2 / 2 \spklrcp^2}}{\sqrt{2 \pi}\pkl \cdot \spklrcp},
\end{eqnarray}
which is highly asymmetric and whose 68\% confidence interval is $[\sigma^{-}_{\pklrc},\sigma^{+}_{\pklrc}]$, where
\begin{equation}\label{spMLmin}
\sigma^{-}_{\pklrc} = {\rm e}^{-\spklrcp} \pklrc,
\end{equation}
\begin{equation}\label{spMLmax}
\sigma^{+}_{\pklrc} =  {\rm e}^{\spklrcp} \pklrc.
\end{equation} 
Furthermore, each $p'_{k0}$ and $p'_{k1}$ are connected by the covariance ${\rm cov}(p'_{k0,{\rm ML}},p'_{k1,{\rm ML}})$, so the confidence intervals of $p_{k0}$ and $p_{k1}$ are not independent. In the next section we explain how to obtain the confidence interval of the galaxy merger fraction taking this into account.

\subsection{The galaxy merger fraction}\label{teofmg}
Expressing the galaxy merger fraction in the range $[z_k, z_{k+1})$ (Equation \ref{fmg}) as a function of the output variables of the ML method we obtain:
\begin{equation}\label{fmgrc}
\ffkml = \frac{2p_{k1,{\rm ML}}}{p_{k0,{\rm ML}}+2p_{k1,{\rm ML}}}.
\end{equation}

However, we cannot obtain the 68\% conficence interval of $\ffkml$, defined as $[\sffkmlm, \sffkmlp]$, applying the usual error theory, which is based in Gaussianity of variables, because the probability distribution of each $\pklrc$ is log-normal. Furthermore, the problem is not analytic and we cannot obtain a mathematical description of the $\ffk$ probability distributions. We made Monte Carlo simulations to characterize the probability distribution of each $\ffk$. The simulations showed that the $\ffk$ distributions can be fit with a log-normal:
\begin{equation}
P_{LN}(\ffk|\ffkml, \sigma) = \frac{{\rm e}^{-(\ln \ffk - \ln \ffkml )^2 / 2 \sigma^2}}{\sqrt{2 \pi} \ffk \cdot \sigma},
\end{equation}
where $\sigma$ is the only free parameter on the fit. Finally, the 68\% confidence interval of $\ffkml$ is given by
\begin{equation}\label{ffsigm}
\sffkmlm = {\rm e}^{-\sigma}\ffkml,
\end{equation}
\begin{equation}\label{ffsigp}
\sffkmlp =  {\rm e}^{\sigma} \ffkml.
\end{equation}

\section{SIMULATIONS WITH SYNTHETIC CATALOGS}\label{simulations}
The accuracy and reliability of the ML method can be tested using synthetic catalogs. This is an important step since ML theory warns that the estimated parameters may suffer from biases; convergence is only assured for large number of sources. The approach is to create catalogs with predefined underlying distribution parameters and compare with the estimated ML parameters. Note that the inputs of the ML method are the same whether we have a real catalog or a synthetic one. In the following paragraphs, we first explain how we created the synthetic catalogs in a general case, and later define and justify the input parameters used for the synthetic catalogs in this paper. Given the high number of variables used in the following discussion, we provide their precise definitions in Table \ref{defvar}.

We created the synthetic catalogs as follows: first we took $n$ sources distributed in redshift and asymmetry space following a bidimensional distribution defined by the input probabilities $\pklr$. This process yielded the $z_{{\rm in},i}$ and $A_{{\rm in},i}$ values of the $n$ sources of our synthetic catalogs, which play the role of $z_{{\rm real},i}$ and $A_{{\rm real},i}$ in Equation \ref{bidi}. Next, we applied the experimental errors: following Equation \ref{zgauss} we obtained the simulated $z_{{\rm sim},i}$ values, which play the role of $z_{{\rm obs},i}$, as drawn from a Gaussian distribution with mean $z_{{\rm in},i}$ and standard deviation $\sigma_{z_{{\rm sim},i}}$; the latter plays the role of $\sigma_{z_{{\rm obs},i}}$. The value of $\sigma_{z_{{\rm sim},i}}$ is a positive value obtained also from a Gaussian distribution with mean $\overline{\sigma_z}$ and standard deviation $\sigma_{\sigma_z}$.  The process was repeated following Equation \ref{agauss} to obtain the simulated $A_{{\rm sim},i}$ and its standard deviation $\sigma_{A_{{\rm sim},i}}$. Finally, we applied the ML method over the synthetic catalog to obtain $\pklrcp$ and $\spklrcp$. Summarizing, the input parameters of our simulations were the bidimensional distribution $\pklr$, $n$, $\overline{\sigma_z}$, $\sigma_{\sigma_z}$, $\overline{\sigma_A}$, and $\sigma_{\sigma_A}$, and the output parameters were $\pklrcp$ and $\spklrcp$.

We defined three intervals in redshift ($k = 0,1,2$) with $\Delta z = 0.4$ and $z \in [0,1.2)$, and two in asymmetry ($l = 0,1$) with $\Delta A = 0.7$ and $A \in [-0.35,1.05)$. Distorted sources with $A > A_{\rm m} = 0.35$ (see C03 for details about the determination of this limit value) are described by $p'_{k1,{\rm in}}$, while normal sources by $p'_{k0,{\rm in}}$. We list in Table \ref{realbidih} the redshift and asymmetry intervals, as well as the probabilities $\pklr$ and $\pklrp = \ln \pklr$, that define the input bidimensional distribution of our synthetic catalogs.  
The $\pklrp$ values in Table \ref{realbidih} do not match any particular observational determination of these quantities, but they follow the general behavior inferred from observed galaxy merger fractions: highly asymmetric galaxies are less frequent than low-asymmetry galaxies up to $z = 1.2$ \citep[][]{conselice03ff, cassata05, bridge07, kamp07}, so the $p'_{k1,{\rm in}}$ are lower than the $p'_{k0,{\rm in}}$. The number of highly asymmetric galaxies increases with redshift in the range $z \in [0,1.2)$ \citep[][]{conselice03ff}, so $p'_{k1,{\rm in}}$ increase with redshift. Several studies present a maximum at intermediate $z$ in the redshift distribution of galaxies in optically selected samples \citep[e.g.,][]{grazian06}, so $p'_{k0,{\rm in}}+p'_{k1,{\rm in}}$ values have a maximum in the interval $z = [0.4,0.8)$. We can check that the $\pklrp$ are normalized following Equation \ref{norm}. Although we preset here this particular bidimensional distribution we carried out the same study with other distributions, and the results were similar.

For convenience we express the experimental  dispersions using the dimensionless variables
\begin{eqnarray}
\sigma_{{\rm bin},z} = \frac{\overline{\sigma_z}}{\Delta z},\label{sbinz}\\
\sigma_{{\rm bin},A} = \frac{\overline{\sigma_A}}{\Delta A}.\label{sbina}
\end{eqnarray}
We used the same value of both variables in each simulation, that is, we used $\sigma_{\rm bin} = \sigma_{{\rm bin},z} = \sigma_{{\rm bin},A}$. Because we fixed the values of $\Delta z = 0.4$ and $\Delta A = 0.7$, $\sigma_{\rm bin}$ unequivocally defines $\overline{\sigma_z}$ and $\overline{\sigma_A}$. It is important to notice that, when we work with observational data, the situation is the opposite: our data define  $\overline{\sigma_z}$ and $\overline{\sigma_A}$, and we should choose the most appropriate values of $\Delta z$ and $\Delta A$. We made simulations for $\sigma_{\rm bin}$ = 0 as a check corresponding to null experimental errors, $\sigma_{\rm bin}$ = 0.25 and 0.5 as typical observational cases, and $\sigma_{\rm bin}$ = 1.0 as extreme case to explore the applicability limits of the ML method. The values of $\sigma_{\sigma_z}$ and $\sigma_{\sigma_A}$ were a half of $\overline{\sigma_z}$ and $\overline{\sigma_A}$ respectively in all cases.

We ran models with $n$ = 50, 100, and 1000 to check catalog size effects. We took these values because we expect experimental catalogs of a few hundred sources or more and we are interested in the applicability limits of the method to small samples.

In order to study how the ML parameters compare with the input parameters, we must preform several simulations and study how the parameters $\pklrcp$ are distributed. Hence, for each $n$ and $\sigma_{\rm bin}$ case we create a simulation set of $N=1000$ independent synthetic catalogs.

\begin{figure*}[t]
\begin{center}
\includegraphics[width = 0.3\linewidth]{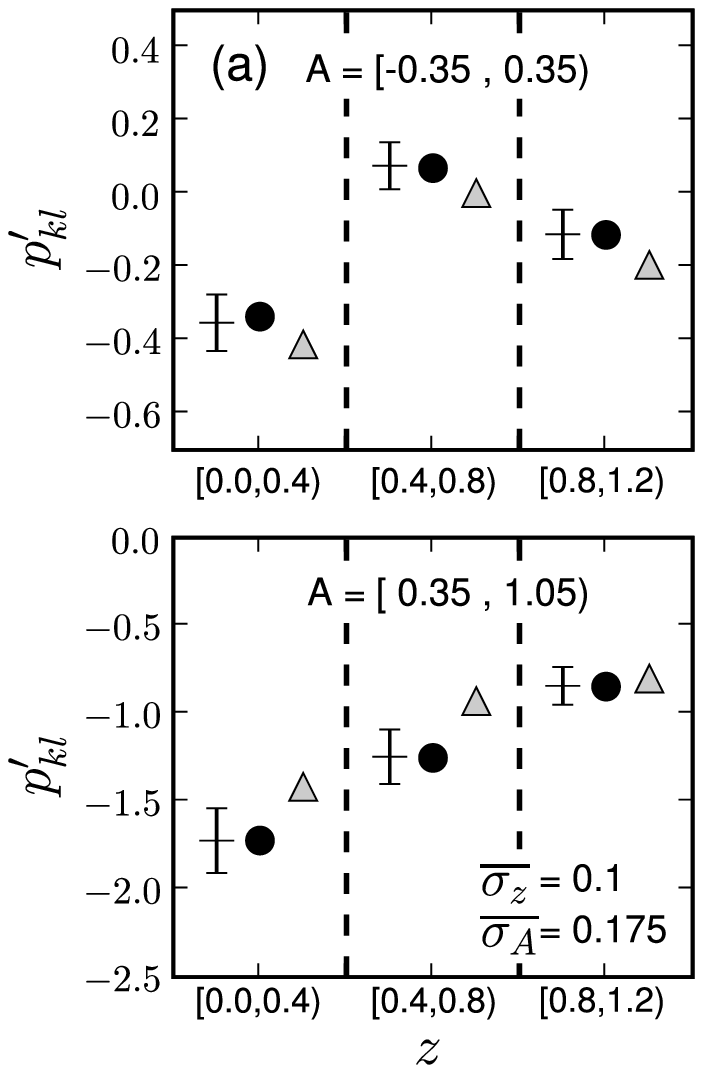}
\includegraphics[width = 0.3\linewidth]{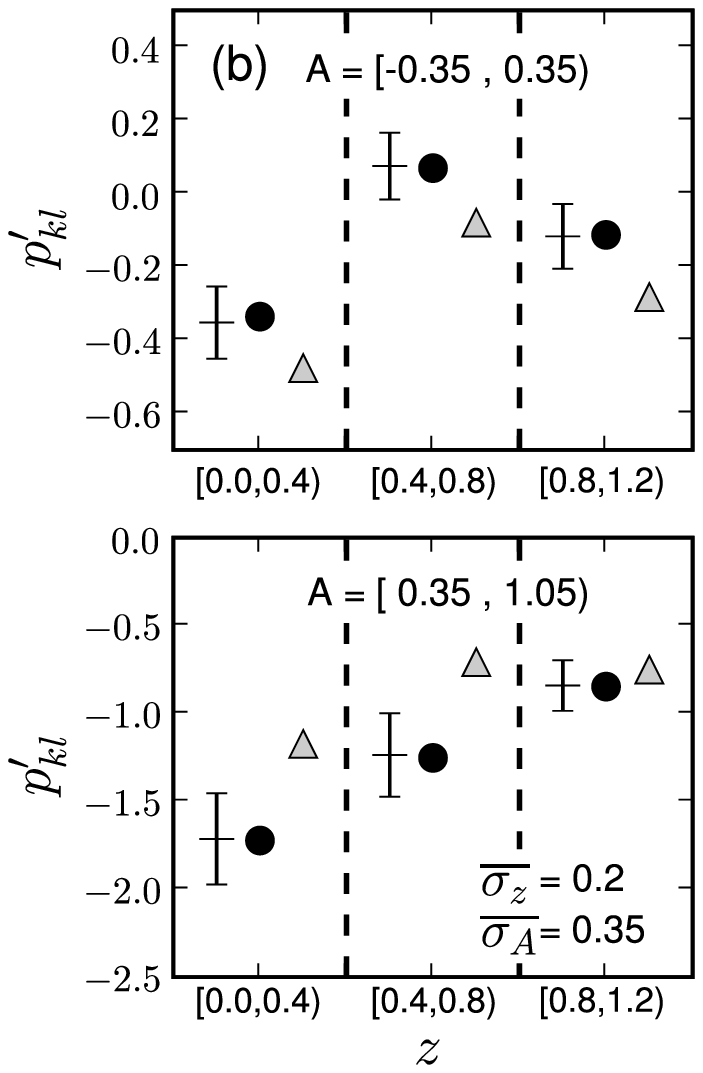}
\includegraphics[width = 0.3\linewidth]{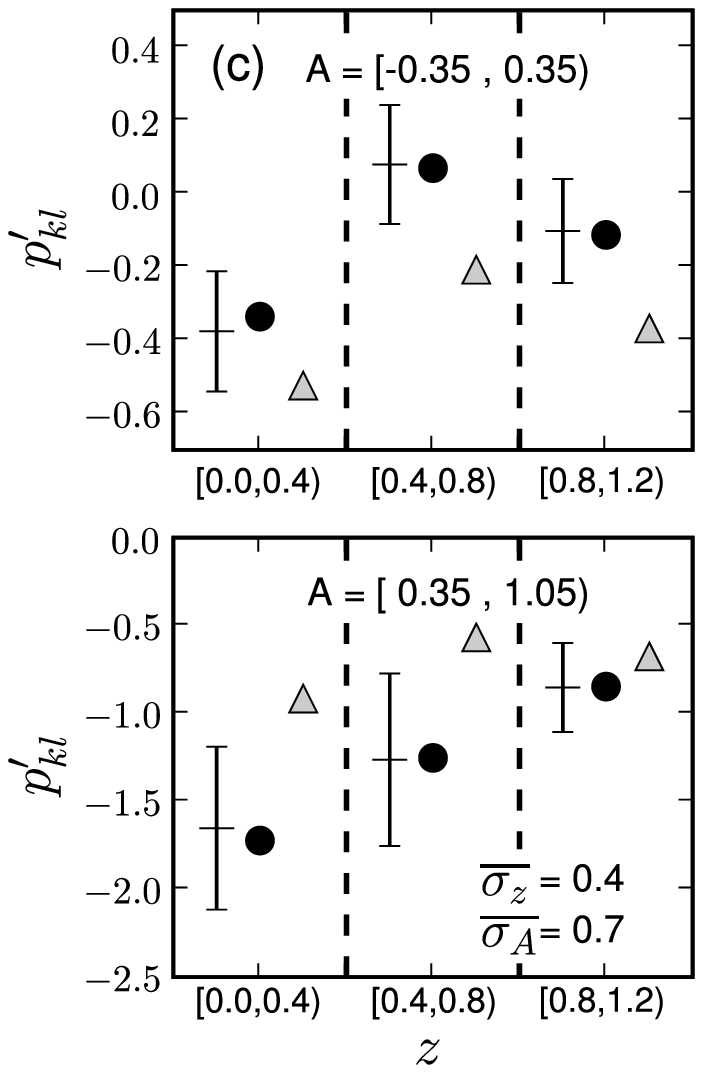}
\caption{Results of run the ML method over $N=1000$ synthetic catalogs with $n=1000$ sources each for different experimental errors: (a) $\sigma_{\rm bin} = 0.25$, (b) $\sigma_{\rm bin} = 0.5$, and (c) $\sigma_{\rm bin} = 1$. In all figures black circles are the input bidimensional probabilities $\pklrp$, gray triangles are the classical bidimensional probabilities $\mpklclassp$ and crosses are the ML bidimensional probabilities $\mpklrcp$. The error bars are the 68\% confidence intervals given by ML method, $[\mpklrcp - \mspklrcp, \mpklrcp + \mspklrcp]$.}
\label{hist1000}
\end{center}
\end{figure*}

The results of the simulations are shown in Figure \ref{hist1000}, and in Tables \ref{simresult50}, \ref{simresult100}, and \ref{simresult1000}. Figure \ref{hist1000} shows $\pklrcp$ for samples of $n=1000$ sources (crosses), with error bars showing their 68\% confidence intervals; for comparison, the input probabilities $\pklrp$ are shown as black circles, and the $\pklclassp$, obtained by drawing a classical histogram (as defined below in Section \ref{class}), are shown as gray triangles, also for $n=1000$ catalogs. In Figure \ref{hist1000}, panels $a$, $b$, and $c$ correspond to increasing values of the experimental errors, defined in Equations \ref{sbinz}, \ref{sbina} and shown in the legend; panels $a,b,c$ may be taken to respectively describe 'good', 'typical', and 'bad' observational errors as compared to the $z$ and $A$ bin sizes. The top/bottom panels show $\pklp$ for the low/high-asymmetry bins. Within each panel, values for the three redshift bins are shown, as labeled 
 on the horizontal axes. We provide the results in tabular format in Tables \ref{simresult50}, \ref{simresult100}, and \ref{simresult1000}, corresponding to simulations with sample sizes of $n=50, 100$, and $1000$, respectively.

\subsection{Classical bidimensional distribution}\label{class}
Before presenting the results of the ML method, we analyze the estimation of the $\pklp$ parameters using the classical bidimensional historgram of the $z_{{\rm sim},i}$ and $A_{{\rm sim},i}$ data. We translate the histogram occupation numbers $n_{kl}$ to probabilities $\pklclassp$ using
\begin{equation}\label{classhist}
\pklclassp = \ln \bigg( \frac{n_{kl}}{\Delta z \Delta A \Sigma_{k} \Sigma_{l} n_{kl} }\bigg),
\end{equation}
where $n_{kl}$ is the number of sources whit $z_{{\rm sim},i}$, $A_{{\rm sim},i}$ whitin the $[z_k, z_{k+1})$ $\cup$ $[A_l, A_{l+1})$ bin. We want to study how the classical method compares with the input parameters. The distribution of the $N$ values of $\pklclassp$ in one simulation set can be
represented by its median $\mpklclassp$ and standard deviation $\smpklclassp$. In Tables \ref{simresult50} - \ref{simresult1000} we can see that the classical bidimensional distribution recovers the input probabilities in the case of null experimental errors and $n$ large as expected. However, the shape of the input bidimensional distribution begins to deviate when $\sigma_{\rm bin}$ increases, as we can also see in Figure \ref{hist1000}: the classical bidimensional distribution (gray triangles) is smoothed by experimental errors and does not estimate well the underlying bidimensional distribution (black circles). We study this in detail in Section \ref{qpkl}.

\subsection{The ML method in absence of experimental errors}
We first test that the ML method, in the case of null experimental errors, recovers the input bidimensional distribution, i.e., that it reduces to the classic method. We can see in Tables \ref{simresult50} - \ref{simresult1000} that the values of $\mpklclassp$ and the median of the $N$ values recovered by the ML method, denoted as $\mpklrcp$, are the same in all cases. This also happens with the values of $\smpklclassp$ and the standard deviations of $\mpklrcp$, denoted as $\smpklrcp$. This indicates that the ML method does not introduce systematic effects on the results.

\begin{figure}[t]
\epsscale{1.0}
\plotone{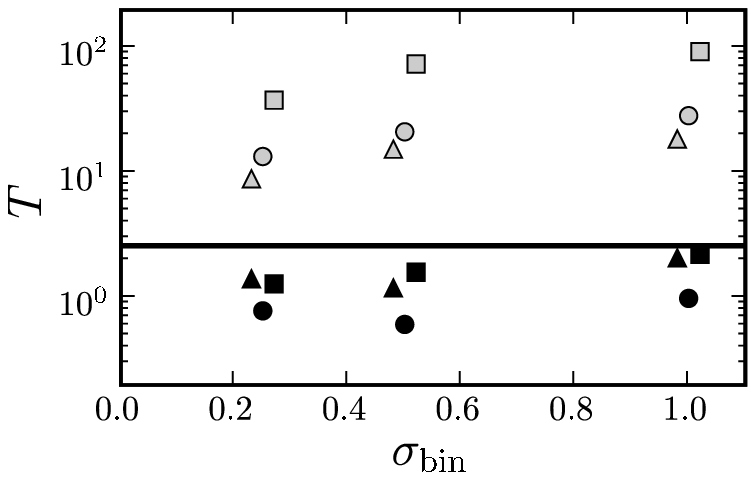}
\caption{Variation of $T_{\rm ML}$ (black symbols) and $T_{\rm class}$ (gray symbols) with dimensionless experimental error size $\sigma_{\rm bin}$. Triangles are for $n = 50$, circles for $n = 100$, and squares for $n = 1000$ source catalogs. The solid line is the 99\% confidence limit $T = 2.6$.}
\label{Tkl}
\end{figure}

\subsection{The ML method with non-null experimental errors}\label{qpkl}
We now examine how well the ML and classical methods recover the input probabilities $\pklrp$ when non-null experimental errors are included in the synthetic catalogs. We use the $N=1000$ source catalogs as an example, which is representative of the general trends. The results are shown in Figure 1, and are tabulated in Table 5. It is clear from Figure 1 that $\pklrcp$  (crosses), recover the input probabilities $\pklrp$ (black circles) in all cases, including those in which the inserted errors are as large as the bin size (panels $c$).  
From Table 5 we see that the values of $\pklrp$ always lay within the 68\% confidence interval of the ML method, defined by $[\mpklrcp - \smpklrcp, \mpklrcp + \smpklrcp]$. This shows that the ML method is reliable. In contrast, the probabilities $\pklclassp$ derived from the classical histogram (gray triangles in Figure 1) systematically deviate from the input probabilities. Probabilities are systematically underestimated/overestimated in the low/high-asymmetry bins (upper/lower panels), due to a spill-over from the most populated bins (low asymmetries) to the least populated, high-asymmetry bins. Such deviations increase for larger experimental errors.  When the errors are as large as the bin size, spill-over is so pronounced that the probabilities in the high-asymmetry sample (lower right panel) are nearly equal for the three redshift bins, and all information on the redshift variation of the galaxy merger fractions is lost.

We conclude that the ML method is an unbiased estimator of the input distribution.  To put this statement in a more quantitative basis, we carry out a Student's t-test \citep[][p. 232]{collins90}. We define our estimator as
\begin{equation}
T_{kl,{\rm ML}} = \frac{\sqrt{N}\left| \pklrp - \mpklrcp \right|}{\smpklrcp},
\end{equation}
and accept that $\pklrp = \mpklrcp$ with a 99\% of confidence when $T_{kl,{\rm ML}} \leq 2.6$. We define in the same way the variable $T_{kl,{\rm class}}$ to study the accuracy of the $\pklclassp$ as an estimator of the $\pklrp$. We calculate the median of the $T_{kl,{\rm ML}}$ and $T_{kl,{\rm class}}$ for each simulation set, denoted as $T_{\rm ML}$ and $T_{\rm class}$ respectively, to make a comparison beetwen different $n$ and $\sigma_{\rm bin}$.

The results are summarized in Tables \ref{simresult50} - \ref{simresult1000}, and in Figure \ref{Tkl}. We can see that $T_{\rm ML}$ is below the confidence level for all $n$ and $\sigma_{\rm bin}$: the $\pklrcp$ are good estimators of the $\pklrp$, as wanted. In contrast, the classical method is far from the confidence condition even in the $\sigma_{\rm bin} = 0.25$ case, and $T_{\rm class}$ increases with $\sigma_{\rm bin}$. Besides, having a large $n$ does not improve the results of classical method: the $\pklclassp$ values are similar for every $n$, but the errors are reduced when increasing $n$, making $T_{\rm class}$ higher. That is, having a large observational sample affected by experimental errors does not improve the estimation of $\pklrp$, and the $\pklclassp$ errors are underestimated. This bias affects the galaxy merger fractions obtained from $\pklclassp$, as we can see on Section \ref{FmgComp}.

\begin{figure}[t]
\epsscale{1.0}
\plotone{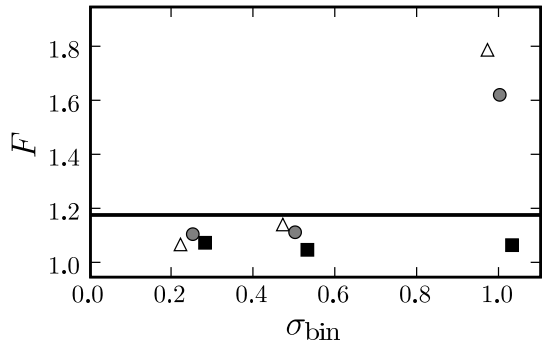}
\caption{Variation of $F$ with dimensionless experimental error size $\sigma_{\rm bin}$. Triangles are for $n = 50$, circles for $n = 100$, and squares for $n = 1000$ source catalogs. The solid line is the 99\% confidence limit $F = 1.8$.}
\label{Fkl}
\end{figure}

\subsection{Study of $\spklp$}\label{qspkl}
When we apply the ML method to an observational sample we obtain an estimation of the $\pklrcp$ 68\% confidence intervals, $[\pklrcp - \spklrcp, \pklrcp + \spklrcp]$, and we want to know if these confidence intervals are representative of the $\pklp$ probability distributions. They are representative if the median of the $N$ values of $\spklrcp$, denoted as $\mspklrcp$, are similar to $\smpklrcp$. To study this issue we perform a Fisher's variance test \citep[][p. 234]{collins90}. We define our estimator as
\begin{equation}
F_{kl} = \frac{\max (\mspklrcp,\smpklrcp)^2}{\min (\mspklrcp,\smpklrcp)^2},
\end{equation}
and accept that $\smpklrcp = \mspklrcp$ with a 99\% of confidence when $F_{kl} \leq 1.18$. We calculate the median of the $F_{kl}$ for each simulation set, denoted as $F$, to make a comparison beetwen different $n$ and $\sigma_{\rm bin}$. The results are summarized in Tables \ref{simresult50} - \ref{simresult1000}, and in Figure \ref{Fkl}. We can see that $\smpklrcp = \mspklrcp$ for all $n$ when $\sigma_{\rm bin} = 0.25, 0.5$. Only when $\sigma_{\rm bin} = 1.0$ and the samples are small ($n=50,100$) does F lie above the confidence limits.

These results imply that the ML method supplies reliable confidence intervals of $\pklrcp$ with thousand sources samples or, with less sources, if the experimental errors are at most a half of the histogram bin size.

The differences between $\smpklrcp$ and $\mspklrcp$ have two origins. The main effect comes from the fact that the probability distributions of $\pklrcp$ are not perfectly Gaussian, and we had assumed Gaussianity to obtain $\spklrcp$ analytically. We study this issue in the next section. The other effect is that we evaluated the theoretical values of $\spklrcp$ at $\pklrcp$: the minimization method \texttt{AMOEBA} is not perfect and we may have estimated a local minimum of Equation \ref{lagrange} instead the absolute minimum (see Section \ref{sigML}).

\subsection{Probability distributions of $\pklp$}\label{gauss}
In the analytical estimation of the $\pklrcp$ covariances we assumed that the $\pklrcp$ probability distributions are Gaussian. To check this assumption we made a histogram of the $N$ values of $\pklrcp$ to obtain the shape of the $\pklrcp$ probability distribution, which we want to approximate by a Gaussian with mean $\mpklrcp$ and standard deviation $\smpklrcp$. We tested this Gaussian approximation with a Kolmogorov-Smirnov test \citep[][p. 235]{collins90}.

We saw that the Gaussian distribution approximation was valid for all $\sigma_{\rm bin}$ in the $n=1000$ simulation sets. The situation of the $n=50$ and 100 simulation sets was more complicated. For $n=100$ the $p'_{k0,{\rm ML}}$ Gaussian approximation was valid for all $\sigma_{\rm bin}$, while the $p'_{k1,{\rm ML}}$ started to be non Gaussian for $\sigma_{\rm bin} = 0.5$, and we could not assume Gaussianity for $\sigma_{\rm bin} = 1.0$. For $n=50$ simulations we could not assume Gaussian approximation from $\sigma_{\rm bin} = 0.25$ to the $p'_{k1,{\rm ML}}$ and from $\sigma_{\rm bin} = 0.5$ to the $p'_{k0,{\rm ML}}$.

These results emphasize that one must check the Gaussian approximation of the $\pklrcp$ probability distributions in each case. That is, when applying the ML method to an experimental catalog it is essential to make special simulations aimed at verifying the Gaussianity of the recovered probabilities.

\subsection{The standard deviation of the ML method due to iterative minimization}\label{sigML}
The iterative minimization method \texttt{AMOEBA} used to obtain the minimum of Equation \ref{lagrange} can introduce an error in the determination of $\pklrcp$ if the method converges to a local minimum. Besides, increasing the experimental errors relaxes the conditions over the absolute minimum and makes it more probable that the method converges onto one such local minimum.
To study this effect and its importance, we apply the ML method $N=100$ times over the same catalog, one per simulation set. We define the variable $\spklMLp$ as the dispersion of the $N$ values of the recovered probabilities $\pklrcp$. We find that the values of $\spklMLp$ depend on the tolerance and the maximum number of iterations of the minimization method. We take a $10^{-15}$ tolerance and 5000 iterations as optimal values: less tolerance or more iterations does not reduce $\spklMLp$, but increased the computational time. All final simulations presented in this paper were made with these optimal values. We also find that $\spklMLp$ increases with $\sigma_{\rm bin}$, but is $\sim 5$ times smaller than $\smpklrcp$ in the worst experimental error case, so the standard deviations of the probabilities are slightly affected by this effect. Therefore, when applying the ML method to an experimental catalog, it is safe practice to apply it more than once, as a precaution against local solutions and iteration bias.

\subsection{The galaxy merger fraction}\label{FmgComp}
In the previous sections we have seen that the experimental errors modify the input bidimensinal distribution, biasing the classical method estimations, whereas the ML method is able to recover the input bidimensional distribution. In this section we study the general effect and trends that the experimental errors introduce on the galaxy merger fraction determination.
To obtain the galaxy merger fraction by the ML method we follow Section \ref{teofmg}. First we determine the galaxy merger fraction $\ffkml$ applying Equation \ref{pML} to the $\pklrcp$ probabilities in Tables \ref{simresult50} - \ref{simresult1000}. Next, we perform Monte Carlo simulations with this $\ffkml$ values and the $\pklrcp$ and $\spklrcp$ in Tables \ref{simresult50} - \ref{simresult1000} to characterize the probability distribution of $\ffk$, obtaining the 68\% confidence interval $[\sffkmlm, \sffkmlp]$ with Equations \ref{ffsigm} and \ref{ffsigp}.

The galaxy merger fraction by the classical method is, applying Equation \ref{fmgf},
\begin{equation}
\ffkclass = \frac{2{\rm e}^{p'_{k1,{\rm class}}}}{{\rm e}^{p'_{k0,{\rm class}}}+2{\rm e}^{p'_{k1,{\rm class}}}}\label{ffclass},
\end{equation}
while its 68\% confidence interval $[\ffkclass - \sffkclass,$ $\ffkclass + \sffkclass]$ is obtained applying the usual error theory to Equation \ref{ffclass},
\begin{eqnarray}\label{sffclass}
\sffkclass = \frac{2{\rm e}^{p'_{k0,{\rm class}}}{\rm e}^{p'_{k1,{\rm class}}}}{({\rm e}^{p'_{k0,{\rm class}}}+2{\rm e}^{p'_{k1,{\rm class}}})^2} \nonumber\\ 
\times \sqrt{s_{p'_{k1,{\rm class}}}^2 + s_{p'_{k0,{\rm class}}}^2}.
\end{eqnarray}

\begin{figure}[t]
\epsscale{1.0}
\plotone{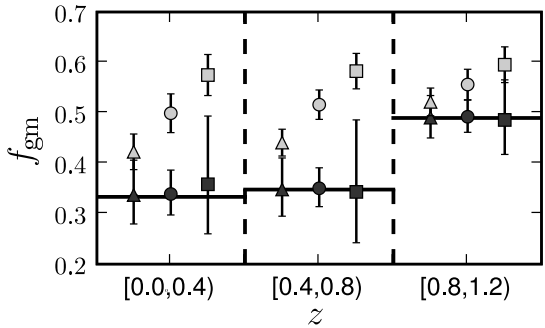}
\caption{Galaxy merger fraction estimations by classical (gray symbols) and ML method (black symbols). In the two cases triangles are for $\sigma_{\rm bin} = 0.25$, circles for $\sigma_{\rm bin} = 0.5$, and squares for $\sigma_{\rm bin} = 1$. The black solid lines are the input galaxy merger fraction in each redshift bin. We can take $\sigma_{\rm bin} = 0.25$ as observational reference.}
\label{ffstudyfig}
\end{figure}

Because of the experimental error limits of the ML method which we noticed in the previous sections, we only made this study with the $n=1000$ simulation sets. We summarize the results in Table \ref{ffstudy}, and Figure \ref{ffstudyfig}. We can see that the classical method gives worst estimates of the input galaxy merger fraction when the experimental errors increase. We may take as observational reference the $\sigma_{\rm bin} = 0.25$ case (for example, in \citealt{conselice03ff} we have $\sigma_{\rm bin} \sim 0.2$). In this case, the difference between the input and the classical estimation is $\sim 0.1$ on the first and second redshift intervals, which have the lower input galaxy merger fraction, and $\sim 0.05$ in the third interval. Furthermore, the experimental errors tend to smooth the galaxy merger fraction values. An extreme case is $\sigma_{\rm bin} = 1$, where the dependency in $z$ has been lost. In addition, the confidence intervals are underestimated and are $\sim 0.035$ in every case. In contrast, the differences between the input and ML method galaxy merger fractions are $\sim 0.01$ in every redshift bin and experimental error case. Furthermore, the 68\% confidence intervals are more realistic: in the $\sigma_{\rm bin} = 0.25, 0.5$ cases they are $\sim 0.05$, while in the $\sigma_{\rm bin} = 1.0$ case they increase to $\sim 0.1$.

Finally, we also determined the classical galaxy merger fraction in the $n=50$ and $100$ cases, and noticed that the values of $\ffkclass$ were similar in each $\sigma_{\rm bin}$ case: having large samples does not improve the results and we must take into account the experimental errors in our analysis to avoid the bias.

\section{DETERMINATION OF ANY ONE- OR BIDIMENSINAL DISTRIBUTION BY THE ML METHOD}\label{MLsteps}

The method outlined here may easily be applied to the unbiased determination of any bidimensional distribution in the presence of observational errors. For example, the automatic indices $M_{20}$ and $G$ are used in \citet{lotz06} to determine the galaxy merger fraction by morphological criteria.  We could apply the ML method by defining the variable $MG = G + 0.14M_{20} - 0.33$ and by calling merger systems all sources with $MG > 0$. Similarly, we may apply the ML method to obtain density of sources in color-color diagrams, especially when we have some condition that separates populations, or to determine one-dimensional histogram of any observational magnitude.

For reference, we provide an outline for the application of the ML method to any one- or bidimensional experimental distribution subject to observational errors:
\begin{enumerate}
\item Define the observational catalog. This catalog cannot be restricted to the interval of interest, e.g., $[z_0, z_k]$, because there are sources both with $z_i < z_0$ and $z_i > z_k$ that could belong to a real bidimensional distribution bin within the range of interest due to the observational errors. In general one should include in the sample those sources with $z_i + 2\sigma_{i} > z_0$ and $z_i - 2\sigma_{i} < z_k$ to avoid incompleteness effects.
\item Apply the ML method to the observational catalog. First, define the bidimensional distribution bins taking into account the size of the observational errors. Next, minimize Equation \ref{lagrange} to obtain the most probable values of $p'_{kl}$, $\pklrcp$. To determine their confidence intervals, calculate the Hessian matrix, Equation \ref{hessian}, with the observational data and the previous $\pklrcp$ values. The diagonal elements of the inverse Hessian matrix provide $\spklrcp$. Notice that we assumed Gaussian experimental errors, Equations \ref{zgauss} and \ref{agauss}, in the development of the ML method. If you need to assume other experimental error distributions, you need to recalculate Equations \ref{lagrange}, \ref{sigmap} and \ref{sigmag} with the new error distributions.
\item Check the results with representative synthetic catalogs. Run simulations with synthetic catalogs to test the accuracy and Gaussianity limits of the method in each particular case following the methodology of sections \ref{qpkl}, \ref{qspkl} and \ref{gauss}. These synthetic catalogs should have the previous $\pklrcp$ as bidimensional distribution input, that is, as $\pklrp$, and similar characteristics to the experimental ones to fix the other input parameters. For example, synthetic and experimental catalogs should have same number of sources $n$, and  $\overline{\sigma_z}$ may be given by the median of the photometric redshift errors in each redshift bin, while, for $\sigma_{\sigma_z}$, one may use the dispersions of these photometric redshift errors. Besides, is important to take into account special cases, e.g., the number of sources with $\zspec$, which have $\sigma_z \sim 0$, in each bin, or avoid unphysical values, e.g., negative redshifts.
\item Determine $p_{kl}$, Equation \ref{pML}, and their confidence intervals, Equations \ref{spMLmin} and \ref{spMLmax}, in the reliable cases.
\end{enumerate}

\section{CONCLUSIONS}\label{conclusions}
We have presented a maximum likelihood method to recover bidimensional distributions of experimental data subject to measurement errors, and applied it to the determination of the galaxy merger fraction based on asymmetry criteria from C03.

The Gaussianity of $\pklrcp$ is the strongest condition on the reliability of the method. From the results, taking into account that typical observational catalogs usually have a few hundred sources, and that the probabilities $p'_{k1}$ would be small, we conclude that the bin of the bidimensional distribution must be at least twice the typical error in redshift in the observational catalog. Within this quality limit, the ML method can recover with accuracy and reliability the lost information due to the experimental errors. Besides, our results have realistic errors with known shapes, which the classical histograms cannot provide.

The ML method presented here may in principle be extended to as many dimensions as required by the astrophysical problem we are addressing. For instance, if we wish to determine variations in the galaxy merger fraction as a function of galaxy mass, errors in the galaxy mass determination would make objects spill over from one mass bin
to the next, biasing the classical histogram approach.  The ML method with an added mass axis would solve the problem.  Even if we are not seeking to determine the variation of the galaxy merger fraction with mass, our parent sample unavoidably has a boundary (e.g., luminosity; mass; color), and observational errors make objects jump in and out of the sample, hence potentially modifying the shape of the distribution we are trying to determine. This extension to higher dimensions is straightforward only when the third variable is independent from the other two. In the
case of a third luminosity or mass axis, this is unfortunately not the case: luminosity and mass depend on galaxy redshift, introducing covariances between the variables. Furthermore, luminosity and mass are affected by incompleteness
functions, making our problem non-analytic. We leave the treatment of this problem for future work.

\acknowledgments
We dedicate this paper to the memory of our six IAC colleagues and friends who met with a fatal accident in Piedra de los Cochinos, Tenerife, in February 2007, with a special thanks to Maurizio Panniello, whose teachings of python were so important for this paper.

This work was supported by the Spanish Programa Nacional de Astronom\'{\i}a y Astrof\'{\i}sica through project number AYA2006-12955.

\clearpage

\begin{deluxetable}{cccccc}
\tabletypesize{\footnotesize}
\tablecolumns{6}
\tablewidth{0pc}
\tablecaption{Input bidimensional distribution used for the synthetic catalogs\label{realbidih}}
\tablehead{
$k$ & $l$ & \pklr & \pklrp & $[z_k, z_{k+1})$ &  $[A_l, A_{l+1})$
}
\startdata
0 & 0 & 0.71428 & -0.33647 & [0, 0.4)   & [-0.35, 0.35)   \\
1 & 0 & 1.07143 &  0.06899 & [0.4, 0.8) & [-0.35, 0.35)   \\
2 & 0 & 0.89286 & -0.11333 & [0.8, 1.2) & [-0.35, 0.35)   \\
\tableline
0 & 1 & 0.17857 & -1.72277 & [0, 0.4)   &  [0.35, 1.05)   \\
1 & 1 & 0.28571 & -1.25276 & [0.4, 0.8) &  [0.35, 1.05)   \\
2 & 1 & 0.42857 & -0.84730 & [0.8, 1.2) &  [0.35, 1.05)   \\   
\enddata
\tablecomments{Variable definitions:\\
$k$: index that scans the redshift bins.\\
$l$: index that scans the asymmetry bins.\\
$\pklr$: probability that a source has redshift in bin $k$ and asymmetry in bin $l$.\\
$\pklrp$: logarithm of $\pklr$.\\
$[z_k, z_{k+1})$: redshift bin $k$.\\
$[A_l, A_{l+1})$: asymmetry bin $l$.\\
}
\end{deluxetable}

\begin{deluxetable}{p{1.5cm} p{12 cm}}
\tabletypesize{\footnotesize}
\tablecolumns{2} 
\tablewidth{0pc}
\tablecaption{Variable definitions for the simulations\label{defvar}}
\tablehead{
Variable & Definition
}
\startdata
\multicolumn{2}{c}{Input Variables}\\[5 pt]
\tableline
$n$ & Number of total sources in a synthetic catalog.\\[5 pt]
$n_{kl}$ & Number of sources in $[z_k, z_{k+1}) \cup [A_l, A_{l+1})$ bin.\\[5 pt]
$\Delta z$ & Redshift bin size.\\[5 pt]
$\Delta A$ & Asymmetry bin size.\\[5 pt]
$N$ & Number of synthetic catalogs in each simulation set.\\[5 pt]
$\pklrp$& Logarithmic probabilities of the input bidimensional distribution of the synthetic catalogs\\[5 pt]
$\overline{\sigma_z}$ & Median experimental errors in redshift of the synthetic catalog sources.\\[5 pt]
$\sigma_{\sigma_z}$ & Dispersion on $\sigma_z$ of the synthetic catalog sources.\\[5 pt]
$\overline{\sigma_A}$ & Median experimental errors in asymmetry of the synthetic catalog sources.\\[5 pt]
$\sigma_{\sigma_A}$ & Dispersion on $\sigma_A$ of the synthetic catalog sources.\\[5 pt]
$\sigma_{\rm bin}$  & $\frac{\overline{\sigma_z}}{\Delta z} = \frac{\overline{\sigma_A}}{\Delta A}$. Dimensionless experimental error size.\\
\cutinhead{Output Variables}
$\pklclassp$&Classical logarithmic probabilities of the classical bidimensional distribution.\\[5 pt]
$\mpklclassp$&Median of the $N$ values of $\pklclassp$ in one simulation set.\\[5 pt]
$\smpklclassp$&Standard deviation of the $N$ values of $\pklclassp$ in one simulation set.\\[5 pt]
$\pklrcp$&Logarithmic probabilities of the bidimensional distribution recovered by the ML method.\\[5 pt]
$\spklrcp$&The 68\% confidence interval of $\pklrcp$ given by the ML method, $[\pklrcp - \spklrcp, \pklrcp + \spklrcp]$.\\[5 pt]
$\mpklrcp$&Median of the $N$ values of $\pklrcp$ in one simulation set.\\[5 pt]
$\smpklrcp$&Standard deviation of the $N$ values of $\pklrcp$ in one simulation set.\\[5 pt]
$\mspklrcp$&Median of the $N$ values of $\spklrcp$ in one simulation set.\\
\cutinhead{Quality Variables}
$T_{kl,{\rm ML}}$ & $\frac{\sqrt{N}\left| \pklrp - \mpklrcp \right|}{\smpklrcp}$. Accepted that $\pklrp = \mpklrcp$ when $T_{kl,{\rm ML}} \leq 2.6$.\\[5 pt]
$F_{kl}$ & $\frac{\max (\mspklrcp,\smpklrcp)^2}{\min (\mspklrcp,\smpklrcp)^2}$. Accepted that $\smpklrcp = \mspklrcp$ when $F_{kl} \leq 1.18$. \\[5 pt]
$\spklMLp$&Standard deviation of the ML method due to iterative minimization process.\\
\enddata
\end{deluxetable}

\begin{deluxetable}{cccccccccc} 
\tabletypesize{\scriptsize}
\tablecolumns{10} 
\tablewidth{0pc} 
\tablecaption{Results of ML method over N = 1000 synthetic catalogs with n = 50 sources\label{simresult50}}
\tablehead{ 
\colhead{\pklp} & \colhead{$\pklrp$} & \colhead{$\mpklrcp$} & \colhead{$T_{kl,{\rm ML}}$} & \colhead{$\smpklrcp$} & \colhead{$\mspklrcp$} & \colhead{$F_{kl}$} & \colhead{$\mpklclassp$} & \colhead{$\smpklclassp$} & \colhead{$T_{kl,{\rm class}}$}\\}
\startdata
\multicolumn{10}{c}{$\overline{\sigma_z} = 0 \ \ \ \sigma_{\sigma_z} = 0 \ \ \ \ \ \ \ \ \overline{\sigma_A} = 0 \ \ \ \sigma_{\sigma_A} = 0 \ \ \ \ \ \ \ \ \sigma_{\rm bin} = 0$}\\[5 pt]
\tableline\\
$p'_{00}$&-0.33647&-0.31627&\nodata&0.30034&0.28284&\nodata&-0.31627&0.30034&\nodata\\
$p'_{10}$&0.06899& 0.08920 &\nodata&0.19871&0.22361&\nodata& 0.08920&0.19871&\nodata\\
$p'_{20}$&-0.11333&-0.13395&\nodata&0.30034&0.25166&\nodata&-0.13395&0.30034&\nodata\\
$p'_{01}$&-1.72277&-1.92571&\nodata&0.81379&0.69282&\nodata&-1.92571&0.81379&\nodata\\
$p'_{11}$&-1.25276&-1.23256&\nodata&0.37839&0.47958&\nodata&-1.23256&0.37839&\nodata\\
$p'_{21}$&-0.84730&-0.82710&\nodata&0.51344&0.38297&\nodata&-0.82710&0.51344&\nodata\\
\cutinhead{$\overline{\sigma_z} = 0.1 \ \ \ \sigma_{\sigma_z} = 0.05 \ \ \ \ \ \ \ \ \overline{\sigma_A} = 0.175 \ \ \ \sigma_{\sigma_A} = 0.0875 \ \ \ \ \ \ \ \ \sigma_{\rm bin} = 0.25$}
$p'_{00}$&-0.33647&-0.33337& 0.29&0.33981&0.34855&1.052&-0.39768&0.34393&  5.63\\
$p'_{10}$& 0.06899& 0.08583& 1.82&0.29197&0.29191&1.001& 0.00779&0.24453&  7.91\\
$p'_{20}$&-0.11333&-0.11093& 0.24&0.31124&0.30611&1.034&-0.20757&0.30034&  9.92\\
$p'_{01}$&-1.72277&-1.78200& 1.82&1.02762&0.87331&1.385&-1.40150&0.53047& 19.15\\
$p'_{11}$&-1.25276&-1.32243& 2.39&0.92324&0.74240&1.546&-0.93512&0.45987& 21.84\\
$p'_{21}$&-0.84730&-0.83122& 1.01&0.50531&0.48405&1.090&-0.82005&0.39750&  2.17\\
\cutinhead{$\overline{\sigma_z} = 0.2 \ \ \ \sigma_{\sigma_z} = 0.1 \ \ \ \ \ \ \ \ \overline{\sigma_A} = 0.35 \ \ \ \sigma_{\sigma_A} = 0.175 \ \ \ \ \ \ \ \ \sigma_{\rm bin} = 0.5$}
$p'_{00}$&-0.33647&-0.34488& 0.59&0.45113&0.47976& 1.131&-0.49339&0.36822& 13.48\\
$p'_{10}$& 0.06899& 0.08788& 1.37&0.43485&0.42913& 1.027&-0.07188&0.29763& 14.97\\
$p'_{20}$&-0.11333&-0.07862& 2.66&0.41200&0.39237& 1.102&-0.29420&0.36126& 15.83\\
$p'_{01}$&-1.72277&-1.88189& 1.05&4.79089&1.45853&10.789&-1.17049&0.55578& 31.42\\
$p'_{11}$&-1.25276&-1.27478& 0.31&2.25784&1.21780& 3.437&-0.72192&0.44502& 37.72\\
$p'_{21}$&-0.84730&-0.87929& 1.34&0.75458&0.70106& 1.158&-0.71523&0.46986&  8.89\\
\cutinhead{$\overline{\sigma_z} = 0.4 \ \ \ \sigma_{\sigma_z} = 0.2 \ \ \ \ \ \ \ \ \overline{\sigma_A} = 0.7 \ \ \ \sigma_{\sigma_A} = 0.35 \ \ \ \ \ \ \ \ \sigma_{\rm bin} = 1.0$}
$p'_{00}$&-0.33647&-0.31312& 1.21&0.60964&0.88500&2.107&-0.53435&0.49716& 12.59\\
$p'_{10}$& 0.06899& 0.15383& 3.79&0.70757&0.85943&1.475&-0.19260&0.39926& 20.72\\
$p'_{20}$&-0.11333&-0.01272& 4.91&0.64776&0.69944&1.166&-0.34675&0.44899& 16.44\\
$p'_{01}$&-1.72277&-2.20397& 1.61&9.46236&5.43006&3.037&-0.93981&0.61357& 40.35\\
$p'_{11}$&-1.25276&-1.78130& 2.55&6.54863&5.83082&1.261&-0.58564&0.55583& 37.95\\
$p'_{21}$&-0.84730&-0.89694& 0.70&2.25333&1.44775&2.422&-0.68095&0.54790&  9.60\\
\enddata 
\end{deluxetable}

\begin{deluxetable}{cccccccccc} 
\tabletypesize{\scriptsize}
\tablecolumns{10} 
\tablewidth{0pc} 
\tablecaption{Results of ML method over N = 1000 synthetic catalogs with n = 100 sources\label{simresult100}}
\tablehead{ 
\colhead{\pklp} & \colhead{$\pklrp$} & \colhead{$\mpklrcp$} & \colhead{$T_{kl,{\rm ML}}$} & \colhead{$\smpklrcp$} & \colhead{$\mspklrcp$} & \colhead{$F_{kl}$} &\colhead{$\mpklclassp$} & \colhead{$\smpklclassp$} & \colhead{$T_{kl,{\rm class}}$}\\}
\startdata
\multicolumn{10}{c}{$\overline{\sigma_z} = 0 \ \ \ \sigma_{\sigma_z} = 0 \ \ \ \ \ \ \ \ \overline{\sigma_A} = 0 \ \ \ \sigma_{\sigma_A} = 0 \ \ \ \ \ \ \ \ \sigma_{\rm bin} = 0$}\\[5 pt]
\tableline\\
$p'_{00}$&-0.33647&-0.33647&\nodata&0.22391&0.20000&\nodata&-0.33647&0.22391&\nodata\\
$p'_{10}$& 0.06899& 0.06899&\nodata&0.14864&0.15275&\nodata& 0.06899&0.14864&\nodata\\
$p'_{20}$&-0.11333&-0.11333&\nodata&0.17864&0.17321&\nodata&-0.11333&0.17864&\nodata\\
$p'_{01}$&-1.72277&-1.72277&\nodata&0.62763&0.43589&\nodata&-1.72277&0.62763&\nodata\\
$p'_{11}$&-1.25276&-1.25276&\nodata&0.37839&0.33912&\nodata&-1.25276&0.37839&\nodata\\
$p'_{21}$&-0.84730&-0.84730&\nodata&0.24924&0.27080&\nodata&-0.84730&0.24924&\nodata\\
\cutinhead{$\overline{\sigma_z} = 0.1 \ \ \ \sigma_{\sigma_z} = 0.05 \ \ \ \ \ \ \ \ \overline{\sigma_A} = 0.175 \ \ \ \sigma_{\sigma_A} = 0.0875 \ \ \ \ \ \ \ \ \sigma_{\rm bin} = 0.25$}
$p'_{00}$&-0.33647&-0.35726& 2.77&0.23715&0.24543&1.071&-0.42902&0.22794& 12.84\\
$p'_{10}$& 0.06899& 0.08249& 2.02&0.21156&0.20096&1.108& 0.01114&0.18577&  9.85\\
$p'_{20}$&-0.11333&-0.11930& 0.81&0.23380&0.21587&1.173&-0.20244&0.20019& 14.08\\
$p'_{01}$&-1.72277&-1.72171& 0.05&0.61932&0.57412&1.164&-1.40984&0.41453& 23.87\\
$p'_{11}$&-1.25276&-1.24035& 0.76&0.51661&0.49041&1.110&-0.92149&0.28759& 36.42\\
$p'_{21}$&-0.84730&-0.84335& 0.37&0.33559&0.34188&1.038&-0.79798&0.28500&  5.47\\
\cutinhead{$\overline{\sigma_z} = 0.2 \ \ \ \sigma_{\sigma_z} = 0.1 \ \ \ \ \ \ \ \ \overline{\sigma_A} = 0.35 \ \ \ \sigma_{\sigma_A} = 0.175 \ \ \ \ \ \ \ \ \sigma_{\rm bin} = 0.5$}
$p'_{00}$&-0.33647&-0.34332& 0.74&0.28963&0.31999&1.221&-0.47807&0.26072& 17.17\\
$p'_{10}$& 0.06899& 0.07331& 0.47&0.28921&0.29111&1.013&-0.07671&0.20306& 22.69\\
$p'_{20}$&-0.11333&-0.11022& 0.35&0.27894&0.27998&1.007&-0.26198&0.23834& 19.72\\
$p'_{01}$&-1.72277&-1.69378& 1.03&0.89067&0.81972&1.181&-1.17948&0.42601& 40.33\\
$p'_{11}$&-1.25276&-1.26124& 0.30&0.88586&0.77025&1.323&-0.72455&0.31834& 52.47\\
$p'_{21}$&-0.84730&-0.85992& 0.88&0.45226&0.46377&1.051&-0.76835&0.32399&  7.71\\
\cutinhead{$\overline{\sigma_z} = 0.4 \ \ \ \sigma_{\sigma_z} = 0.2 \ \ \ \ \ \ \ \ \overline{\sigma_A} = 0.7 \ \ \ \sigma_{\sigma_A} = 0.35 \ \ \ \ \ \ \ \ \sigma_{\rm bin} = 1.0$}
$p'_{00}$&-0.33647&-0.34227& 0.40&0.45903&0.57113&1.548&-0.52269&0.33388& 17.64\\
$p'_{10}$& 0.06899& 0.10014& 2.25&0.43673&0.56965&1.701&-0.18589&0.26832& 30.04\\
$p'_{20}$&-0.11333&-0.08903& 1.63&0.46981&0.47632&1.028&-0.39434&0.32866& 27.04\\
$p'_{01}$&-1.72277&-1.82510& 0.65&4.99004&2.17650&5.256&-0.91894&0.40459& 62.83\\
$p'_{11}$&-1.25276&-1.38142& 1.32&3.07524&2.24256&1.880&-0.58439&0.34776& 60.78\\
$p'_{21}$&-0.84730&-0.84699& 0.01&0.85518&0.87014&1.035&-0.65849&0.35630& 16.76\\
\enddata 
\end{deluxetable}

\begin{deluxetable}{cccccccccc} 
\tabletypesize{\scriptsize}
\tablecolumns{10} 
\tablewidth{0pc} 
\tablecaption{Results of ML method over N = 1000 synthetic catalogs with n = 1000 sources\label{simresult1000}}
\tablehead{ 
\colhead{\pklp} & \colhead{$\pklrp$} & \colhead{$\mpklrcp$} & \colhead{$T_{kl,{\rm ML}}$} & \colhead{$\smpklrcp$} & \colhead{$\mspklrcp$} & \colhead{$F_{kl}$} &\colhead{$\mpklclassp$} & \colhead{$\smpklclassp$} & \colhead{$T_{kl,{\rm class}}$}\\}
\startdata
\multicolumn{10}{c}{$\overline{\sigma_z} = 0 \ \ \ \sigma_{\sigma_z} = 0 \ \ \ \ \ \ \ \ \overline{\sigma_A} = 0 \ \ \ \sigma_{\sigma_A} = 0 \ \ \ \ \ \ \ \ \sigma_{\rm bin} = 0$}\\[5 pt]
\tableline\\
$p'_{00}$&-0.33647&-0.33647&\nodata&0.06137&0.06325&\nodata&-0.33647&0.06137&\nodata\\
$p'_{10}$& 0.06899& 0.06899&\nodata&0.04940&0.04830&\nodata& 0.06899&0.04940&\nodata\\
$p'_{20}$&-0.11333&-0.11333&\nodata&0.05336&0.05477&\nodata&-0.11333&0.05336&\nodata\\
$p'_{01}$&-1.72277&-1.72277&\nodata&0.14864&0.13784&\nodata&-1.72277&0.14864&\nodata\\
$p'_{11}$&-1.25276&-1.25276&\nodata&0.11132&0.10724&\nodata&-1.25276&0.11132&\nodata\\
$p'_{21}$&-0.84730&-0.84730&\nodata&0.08066&0.08563&\nodata&-0.84730&0.08066&\nodata\\
\cutinhead{$\overline{\sigma_z} = 0.1 \ \ \ \sigma_{\sigma_z} = 0.05 \ \ \ \ \ \ \ \ \overline{\sigma_A} = 0.175 \ \ \ \sigma_{\sigma_A} = 0.0875 \ \ \ \ \ \ \ \ \sigma_{\rm bin} = 0.25$}
$p'_{00}$&-0.33647&-0.35327& 6.99&0.07596&0.07700&1.028&-0.41246&0.07170&  33.51\\
$p'_{10}$& 0.06899& 0.07525& 3.19&0.06204&0.06414&1.139& 0.00102&0.05671&  37.90\\
$p'_{20}$&-0.11333&-0.11170& 0.73&0.07046&0.06730&1.094&-0.19532&0.06814&  38.05\\
$p'_{01}$&-1.72277&-1.72333& 0.10&0.18378&0.18398&1.005&-1.41729&0.12568&  76.86\\
$p'_{11}$&-1.25276&-1.24578& 1.35&0.16351&0.15516&1.138&-0.92943&0.09163& 111.59\\
$p'_{21}$&-0.84730&-0.84306& 1.23&0.10901&0.10726&1.060&-0.80147&0.08635&  16.78\\
\cutinhead{$\overline{\sigma_z} = 0.2 \ \ \ \sigma_{\sigma_z} = 0.1 \ \ \ \ \ \ \ \ \overline{\sigma_A} = 0.35 \ \ \ \sigma_{\sigma_A} = 0.175 \ \ \ \ \ \ \ \ \sigma_{\rm bin} = 0.5$}
$p'_{00}$&-0.33647&-0.35273& 5.28&0.09732&0.09866&1.027&-0.47684&0.08066&  55.03\\
$p'_{10}$& 0.06899& 0.07463& 1.83&0.09727&0.09114&1.069&-0.07992&0.06535&  72.06\\
$p'_{20}$&-0.11333&-0.11731& 1.36&0.09237&0.08829&1.096&-0.28307&0.07064&  75.99\\
$p'_{01}$&-1.72277&-1.71352& 1.12&0.26019&0.25957&1.002&-1.17318&0.13081& 132.86\\
$p'_{11}$&-1.25276&-1.23600& 2.09&0.25323&0.23734&1.110&-0.70824&0.09678& 177.92\\
$p'_{21}$&-0.84730&-0.84142& 1.25&0.14808&0.14385&1.033&-0.75047&0.10000&  30.62\\
\cutinhead{$\overline{\sigma_z} = 0.4 \ \ \ \sigma_{\sigma_z} = 0.2 \ \ \ \ \ \ \ \ \overline{\sigma_A} = 0.7 \ \ \ \sigma_{\sigma_A} = 0.35 \ \ \ \ \ \ \ \ \sigma_{\rm bin} = 1.0$}
$p'_{00}$&-0.33647&-0.37667& 8.27&0.15372&0.16437&1.143&-0.52454&0.10316&  57.65\\
$p'_{10}$& 0.06899& 0.07901& 1.95&0.16218&0.16229&1.001&-0.20842&0.08748& 100.28\\
$p'_{20}$&-0.11333&-0.10270& 2.45&0.13490&0.14196&1.107&-0.36879&0.09448&  85.51\\
$p'_{01}$&-1.72277&-1.65260& 5.03&0.44100&0.46316&1.103&-0.91542&0.13080& 195.18\\
$p'_{11}$&-1.25276&-1.26323& 0.67&0.49175&0.49023&1.006&-0.56759&0.11510& 188.24\\
$p'_{21}$&-0.84730&-0.85242& 0.65&0.24892&0.25306&1.033&-0.67546&0.11284&  48.15\\
\enddata 
\end{deluxetable}

\begin{deluxetable}{ccccccccc} 
\tabletypesize{\scriptsize}
\tabletypesize{\footnotesize}
\tablecolumns{9}
\tablewidth{0pc}
\tablecaption{Real, ML method, and classic galaxy merger fraction\label{ffstudy}}
\tablehead{
$[z_k, z_{k+1})$ & \ffkr & \multicolumn{3}{c}{\ffkml} &  &\multicolumn{3}{c}{\ffkclass}\\\cline{3-5}\cline{7-9}
 & & $\sigma_{\rm bin} = 0.25$ & $\sigma_{\rm bin} = 0.5$ & $\sigma_{\rm bin} = 1.0$ & &$\sigma_{\rm bin} = 0.25$ & $\sigma_{\rm bin} = 0.5$ & $\sigma_{\rm bin} = 1.0$ 
}
\startdata
$[0, 0.4)$   & 0.3333 & $0.337^{+0.069}_{-0.057}$ & $0.339^{+0.048}_{-0.042}$ & $0.358^{+0.134}_{-0.098}$& &$0.423\pm0.035$ &$0.499\pm0.038$ &$0.575\pm0.040$ \\
$[0.4, 0.8)$ & 0.3478 & $0.348^{+0.063}_{-0.053}$ & $0.350^{+0.040}_{-0.036}$ & $0.343^{+0.140}_{-0.099}$& &$0.441\pm0.027$ &$0.516\pm0.029$ &$0.583\pm0.035$ \\
$[0.8, 1.2)$ & 0.4897 & $0.490^{+0.044}_{-0.040}$ & $0.492^{+0.032}_{-0.030}$ & $0.486^{+0.079}_{-0.068}$& &$0.522\pm0.027$ &$0.556\pm0.030$ &$0.595\pm0.035$ \\  
\enddata
\end{deluxetable}

\end{document}